\documentclass[a4paper, 11pt]{article}

\usepackage{amsfonts, amsmath, amssymb, amsthm, amscd, setspace, subfigure}
\usepackage[final]{graphicx}
\usepackage[usenames,dvipsnames]{color}
\usepackage{jheppub}
\newcommand{\m}{\eta}

\newcommand{\norm}[1]{\ensuremath{\left |\left | #1 \right | \right | }}

\makeatletter
\def\@fpheader{\relax}
\makeatother

\title{Gravitational instantons as models for charged particle systems}
\author{Guido Franchetti}
\author{and Nicholas S.~Manton}
\affiliation{DAMTP, Centre for Mathematical Sciences,\\
University of Cambridge,\\
Wilberforce Road, Cambridge CB3 0WA, UK}

\emailAdd{G.Franchetti@damtp.cam.ac.uk}
\emailAdd{N.S.Manton@damtp.cam.ac.uk}

\abstract{In this paper we propose ALF gravitational instantons of types $A _k $ and $D _k $ as models for charged particle systems.
We calculate the charges of the two families. These are $-(k +1) $ for $A _k $, which is proposed as a model for $k+1 $ electrons, and $2-k $ for $D _k $, which is proposed as a model for either a particle of charge $+2$ and $k$ electrons or a proton and $k-1 $ electrons. Making use of preferred topological and metrical structures of the manifolds, namely metrically preferred representatives of middle dimension homology classes, we construct two different  energy functionals which reproduce the Coulomb interaction energy for a system of charged particles.}


\arxivnumber{}
\begin{document}
\maketitle
\flushbottom
\section{Introduction}
\label{intro} 
In a recent paper \cite{Atiyah:2012hw}, Atiyah, Manton and Schroers have proposed describing elementary particles in terms of 4-dimensional Riemannian manifolds having self-dual Weyl  tensor. Electrically neutral particles are described by compact manifolds, while charged ones are described by non-compact manifolds which have an asymptotic circle fibration over $\mathbb{R}   ^3 $ or its quotient by a finite group. The class of non-compact manifolds considered includes some types of gravitational instanton, originally introduced by Hawking \cite{Hawking:1976jb} as non-singular 4-dimensional solutions of the Euclidean Einstein equations having a Riemann tensor which decays at infinity.  Conserved quantum numbers are identified with topological properties of the manifold: charge is given by minus the first Chern number of the asymptotic fibration, baryon number by signature --- but this last identification was provisional. These ideas were inspired by higher dimensional constructions  related to the (apparently) 3-dimensional physics of our world, like  the Kaluza-Klein circle dimension, and the derivation of approximate Skyrmions from self-dual Yang-Mills fields, i.e.~4-dimensional instantons \cite{Atiyah:1989th}.

In \cite{Atiyah:2012hw}, concrete proposals were made for the neutron, given by $\mathbb{C}  P ^2 $ with the Fubini-Study metric, the proton, given by the Atiyah-Hitchin (AH) manifold \cite{Atiyah:wf},  the electron, given by the Taub-NUT (TN) manifold \cite{Hawking:1976jb}.\footnote{In keeping with the notation used in \cite{Atiyah:2012hw}, we call Atiyah-Hitchin manifold the simply connected double cover of the moduli space of centered $SU(2) $-monopoles of charge $2$. }
 The models are purely static: time and dynamics are not included. The purpose of this paper is twofold: first, two infinite families of manifolds, namely gravitational instantons of types $A _k $ and $D _k $, are presented as candidates for systems of charged particles. The metric tensors of these manifolds depend on a family of parameters which can be interpreted as the classical positions of the particles. The $A _k $ family is proposed as a model for  $k + 1$ electrons; the $D _k  $ family can describe either a particle of charge $+2$ and $k$ electrons or a proton and $k-1 $ electrons.  Second, two expressions for the total energy of the system are presented. When applied to the $A _k $ and $D _k $ families, they reproduce the Coulomb interaction energy of the particles.
Since we believe that $A _k $ and $D _k $  model systems of particles with at most one baryon, the Coulomb energy represents, after the  rest mass, the main contribution to the energy of the system. A rest mass term  is also obtained in one of the two approaches.

The current definition of a (non-compact)  gravitational instanton, see e.g. \cite{Minerbe:2010wta}, is that of a hyperk\"ahler 4-manifold with curvature tensor decaying at infinity. A gravitational instanton is then Ricci flat and has self-dual Weyl and Riemann tensors. The definition is different from the original one given by Hawking, which would include solutions like Euclidean Schwarzschild, for which the Riemann tensor is not self-dual.  Gravitational instantons are classified into four families \cite{Etesi:2006bc}, ALE, ALF, ALG, ALH, which have different volume growth at infinity \cite{Minerbe:2010wta}, e.g.~as $r ^4 $ in the ALE case and as $r ^3 $ in the ALF case. Since non-compact manifolds modelling charged particles are required to approach at infinity an $S ^1 $ fibration over $ \mathbb{R}  ^3 $ (or its quotient by a finite group), we must select the ALF case. It is conjectured \cite{Cherkis:1999gx} that the ALF family is exhausted by two infinite families of manifolds: of type $ A _k $ and of type $ D _k $. At the topological level, the $ A _k $, $k \geq 1 $ and $ D _k $, $k \geq 3 $, manifolds arise as the minimal resolution of Kleinian singularities of cyclic and dihedral type. They retract onto a configuration of 2-spheres intersecting according to the Cartan matrix of the corresponding Lie algebra. Outside a compact set, the topology is that of $\mathbb{C}  ^2 / \Gamma $ where $\Gamma = \mathbb{Z}  _{ k + 1 }$ for $A _k $, $k \geq 1 $, and $\Gamma =D ^\ast _{ k-2 }$, the binary dihedral group of order $4(k-2 ) $, for $D _k $, $k \geq 3 $. Small values of $k$ need a separate description: $A _0 $, the TN manifold, has the topology of $\mathbb{R}   ^4 $; $D _0 $ retracts onto $\mathbb{R}  P ^2 $ and  is not simply connected; $D _1 $ retracts onto $S ^2 $ and outside a compact set has the topology of $\mathbb{C}  ^2 $ quotiented by the $\mathbb{Z}  _4 $ action generated by $( z _1 , z _2 )\mapsto i(- \bar z _2 , \bar z _1 )$; $D _2 $ is the minimal resolution of singularities of $(\mathbb{R} ^3  \times S ^1)/ \mathbb{Z}  _2 $. The $A _k $ and $D _k $ families include the proposed models for the electron,  $A _0 $, and the proton, described by the AH manifold, a member of the $ D _1 $ family.
 
The metric of the $A _k $ family, also known as multi TN \cite{Gibbons:1979xm},  is explicitly known:
\begin{equation}
\label{akmetric} 
\mathrm{d} s ^2 = V \left( \mathrm{d} r ^2 + r ^2 \mathrm{d} \Omega ^2 \right) + V ^{-1} \left( \mathrm{d} \psi  + \alpha  \right) ^2
\end{equation}
with 
\begin{equation}
\label{vak} 
V =1+\sum _{ i =1 }^{k+1} \frac{\m}{\norm{p-p _i}},
\end{equation} 
where $\m>0 $. Here $r\in[0, \infty )$, $\theta\in[0, \pi  ]$, $\phi\in[0, 2 \pi )$ are spherical coordinates, $ \mathrm{d} \Omega ^2 =\mathrm{d} \theta ^2 + \sin ^2 \theta\, \mathrm{d} \phi ^2$ and $\psi $ is an angle having the range $\psi \in[0,4 \pi \m ) $, chosen so as to avoid conical singularities in the metric. The point $p$ has spherical coordinates $(r, \theta, \phi)$ and $\{p _i\}$ are the positions of $k+1$  distinct points in $\mathbb{R}  ^3 $, the NUTs. 
If  $(x, y, z)$ are the Cartesian coordinates of the point $p$,  related to the spherical coordinates $(r, \theta, \phi)$ in the usual way, and $(x _i, y _i, z _i )$ are the Cartesian coordinates of  $p _i $, then $\norm{p-p _i }=\sqrt{ (x-x _i )^2 + (y - y _i )^2 + (z - z _i )^2 } $. The metric  $\mathrm{d} r ^2 + r ^2 \mathrm{d} \Omega ^2$ is the flat metric on $\mathbb{R} ^3 $. The one form $\alpha$ is locally such that $\mathrm{d} \alpha =* _3 \mathrm{d} V $, where $* _3 $ denotes the Hodge star with respect to this flat metric. 
Geometrically, a NUT is a fixed point of the action of the Killing vector $\partial / \partial \psi $ on the manifold \cite{Gibbons:1979xm}.

The metric of the $ D _k $ family is, with the exception of $ D _0 $ and $ D _1 $, known only implicitly \cite{Cherkis:2005ba}; the $D _2 $ metric is presented as an approximation to the K3 metric in \cite{Hitchin:1984wt}. However, for large $r$ the following  asymptotic approximation is known \cite{Chalmers:1999vb}:
\begin{equation}
\label{dkmetric} 
\mathrm{d} s ^2 =V \left( \mathrm{d} r ^2 + r ^2 \mathrm{d} \Omega ^2 \right) + V ^{-1} \left( \mathrm{d} \psi + \alpha \right) ^2 
\end{equation}  
with 
\begin{equation}
\label{vdk} 
V =1- \frac{2\m}{\norm {p }} + \sum _{ i =1 }^k \frac{\m}{2} \left( \frac{1}{\norm {p-p _i }}+ \frac{1}{\norm{p + p _i}} \right),
\end{equation} 
where  $\{\pm p_i \}$ are $2k$ distinct points in $\mathbb{R}  ^3 $, $\mathrm{d} \alpha =* _3 \mathrm{d} V $, $\theta \in [0, \pi ]$, $\phi \in [0, 2 \pi )$, $ \psi \in [0, 2 \pi \m) $,  and there is the additional $\mathbb{Z}  _2 $ identification
\begin{equation}
\label{z2} 
(\theta, \phi, \psi ) \sim (\pi - \theta, \phi + \pi ,- \psi ).
\end{equation} 
For $r \rightarrow \infty $ the positions of the NUTs become irrelevant and the metric has the leading asymptotic form (\ref{dkmetric}) with $V$ given by 
\begin{equation}
\label{vdkas} 
V=1+\frac{(k-2)\m}{r},
\end{equation}
where $r =\norm{p} $. Note that (\ref{vdk}) is given by the superposition of a NUT of negative weight located at the origin and $2k $ NUTs at
$\{ \pm p _i  \} $, and is  symmetric under inversion in the origin. Close to each NUT $p _i $, the metric is  well approximated by the TN metric. If the NUTs are very distant from the origin, $\norm {p _i }\gg 1 $, then eq.~(\ref{dkmetric}), with $V$ given by eq.~(\ref{vdk}), reduces to the asymptotic form of the $D _0 $ metric. From the moduli space description of $ D _k $, see \cite{Cherkis:1999gx,Cherkis:1998cz}, it follows that it is possible to set $  p _k  =0$  without making the manifold singular or altering its topology. Doing so, eq.~(\ref{vdk}) becomes
\begin{equation}
\label{vdkbis} 
V =1- \frac{\m}{\norm {p }} + \sum _{ i =1 }^{k-1} \frac{\m}{2} \left( \frac{1}{\norm {p-p _i }}+ \frac{1}{\norm{p + p _i}} \right).
\end{equation} 
If the other $2 (k-1 )$ NUTs are far from the origin, eq.~(\ref{dkmetric}), with $V$ given by eq.~(\ref{vdkbis}), reduces to the asymptotic form of the AH metric. The  $D _1 $ family, whose asymptotic metric is parametrised by $p _1 $, has been studied by Dancer \cite{Dancer:1994tn}; for $p _1 =0 $ one obtains the AH manifold.

\section{Charge}
\label{charge} 
The kinds of particle systems that can be modelled by $A _k $ and $D _k $ are strongly constrained by the electric charges of the manifolds, which we now calculate. All the manifolds that we consider have an asymptotic $U(1) $-fibration over a base $B$. Charge $Q$  is defined to be minus the first Chern number $c _1 $ of this asymptotic fibration, the integer
\begin{equation}
\label{c1} 
Q =- c _1 =\frac{1}{2 \pi } \int _{B}  F,
\end{equation} 
where $F$ is the 2-form field strength of a connection on the $U(1)$-bundle over $B$.

For $A _k $,
$B =S ^2 $ and the total space of the fibration is $S ^3 / \mathbb{Z}  _{ k + 1 }$ if $k\geq 1 $, $S ^3 $ if $k =0 $. For $D _k $,  because of the identification (\ref{z2}),
  $B =\mathbb{R}  P ^2 $. If $k\geq 3 $, the total space is $S ^3 / D ^\ast _{ k-2 }$; if $k =2 $, the circle bundle over $\mathbb{R}  P ^2 $ is trivial;
if $k = 1 $, the asymptotic topology is the same as that of  $D _3 $, but with opposite orientation. In all cases  $c _1  $ can be computed by considering a double cover $\overline{D}_k  $ of $D _k $ where the identification (\ref{z2}) is lifted, so that $B =S ^2 $. Then  $c _1$ is half the first Chern number of the asymptotic fibration in $\overline{D}_k $.

It is convenient to consider the following connection form, defined on a $U (1) $-bundle over $S ^2 $, which appears in the asymptotic metrics of both $A _k $ and $D _k $,
\begin{equation}
\label{conng} 
\omega _{ l } =
\begin{cases}
\mathrm{d} \psi _N + \left( A _{l} \right)_N & \text{on $U _N $}\\
\mathrm{d} \psi _S + \left( A _{l} \right)_S & \text{on $U _S $}
\end{cases},
\end{equation} 
where $U _N= S ^2 \backslash \{(0,0,-1) \} $, $U _S = S ^2 \backslash \{(0,0,1)\}$, $\psi _N$, $\psi _S$ $\in [0, 2 \pi )$, $\psi _S = \psi _N - l \phi $. The gauge potentials $  \left( A _{l} \right) _N$, $ \left( A _{l} \right)_S$ are given by
\begin{gather}  
\label{dirac} 
\left. \begin{array}{lr} 
 \left( A _{l} \right)   _N
 \\
 \left( A _{l} \right)   _S
 \end{array} \right \} 
 =\frac{l}{2}  \left(  \cos \theta  \mp 1\right)  \mathrm{d} \phi,
\end{gather}
with  $(A _{l} )   _N $ singular for $\theta =  \pi  $, $(A _{l} )   _S$ singular for $ \theta = 0  $.
The  connection form $\omega _l $ is instead globally defined on the total space of the fibration. For $l =0 $, $\omega _0 $ is a flat connection on the trivial bundle $S ^2 \times U(1) $.
Since $U (1 )$ is abelian, $ F _{ l } = \mathrm{d}  \left( A _{l} \right)   _S = \mathrm{d}  \left( A _{l} \right)   _N = - (l/2 )\sin \theta\, \mathrm{d} \theta \wedge \mathrm{d} \phi $
 is globally defined on $S ^2 $.
The first Chern number of a $U (1) $-fibration over $S ^2 $ with connection form $\omega _l $ is $l$, in fact
\begin{equation}
c _1 =
 - \frac{1}{2 \pi } \int _{ S ^2 } F _l =
\frac{l}{4 \pi } \int _{ S ^2 } \sin \theta\, \mathrm{d} \theta \wedge \mathrm{d} \phi =
l.
\end{equation} 
For both $A _k $ and $D _k $ we can identify $l$ by rewriting the asymptotic form of the metric in a way that explicitly shows $\omega _l $.
For $A _k $, asymptotically (\ref{vak}) becomes $V \sim 1 + (k+1)\m/r $, therefore locally $ \alpha  \sim (k + 1 ) \m \cos  \theta \,  \mathrm{d} \phi $. 
Hence, rescaling $ \tilde r =r/\m $, $\tilde s =s/\m$, $ \psi _N  =\psi / (2\m) + (k + 1) \phi/2\in [0, 2\pi)$, $ \psi _S =\psi /(2\m)  -  (k + 1) \phi/2 \in [0, 2\pi) $, the metric (\ref{akmetric}) asymptotically can be written
\begin{equation}
\label{akmetricas} 
\begin{split} 
\mathrm{d}  \tilde s ^2 &\sim
\begin{cases}
  V( \mathrm{d} \tilde r ^2 + \tilde r ^2 \mathrm{d} \Omega ^2 ) + 4V ^{-1} \Big (
 \mathrm{d}  \psi _N + \frac{k + 1 }{2}(  \cos \theta -1 ) \mathrm{d} \phi\Big ) ^2   &\text{for $ \theta , \phi \in U _N $ }\\
   V( \mathrm{d} \tilde r ^2 + \tilde r ^2 \mathrm{d} \Omega ^2 ) + 4V ^{-1}\Big ( 
 \mathrm{d}  \psi _S +\frac{k + 1}{2}( \cos \theta  + 1 ) \mathrm{d} \phi \Big ) ^2  &\text{for $ \theta , \phi \in U _S $ }  
\end{cases}
\\ &=  V( \mathrm{d} \tilde r ^2 + \tilde r ^2 \mathrm{d} \Omega ^2 ) + 4V ^{-1} 
 \left( \omega _{ k + 1 } \right) ^2.
 \end{split} 
\end{equation}
Therefore $l =k + 1 $ and the charge $Q _{ A _k }$ of $A _k $ is
 \begin{equation}
\label{qak} 
Q_{ A _k } = - c _1 
=-(k + 1 ).
\end{equation}

Consider now $D _k $ and its charge $Q _{ D _k }$.
Using (\ref{vdkas}), one has $\alpha \sim (k-2 ) \m\cos \theta \, \mathrm{d} \phi $. 
This time, making the rescaling $\tilde r = r/\m$, $\tilde s = s/\m$,
$ \psi _N  =\psi / \m + (k -2) \phi\in [0, 2\pi)$,
 $ \psi _S  =\psi / \m - (k -2) \phi\in [0, 2\pi)$,
 we find that the leading asymptotic form of the metric (1.3) can be written
\begin{equation}
\mathrm{d}  \tilde s ^2 \sim  V( \mathrm{d} \tilde r ^2 + \tilde r ^2 \mathrm{d} \Omega ^2 ) + V ^{-1} \left( \omega _{ 2 (k - 2 ) } \right) ^2 ,
\end{equation} 
with $V$ given by (\ref{vdkas}). Therefore $l =2 (k -2 )$. Dividing by 2 to pass from $\overline{D}_k $ to $D _k $, we get the charge
\begin{equation} 
\label{qdk} 
Q_{ D _k } =
  -\frac{1}{2} \cdot  2 (k -2 ) = 2-k.
\end{equation}
%

The charge (\ref{qak}) of  $A _k $ is $k+1 $ times that of an electron ($A _0 $) and suggests to  consider  $A _k $ as a model for $k + 1 $ electrons. This interpretation is supported by the form of the metric (\ref{akmetric}), which looks like $A _0 $ close to each NUT. The charge (\ref{qdk}) of  $D _k $ agrees both with that of a proton and $k-1 $ electrons and with that of a particle of charge $+ 2 $ and $k$ electrons.  Indeed, $D _k $ can describe both. If none of the NUTs is at the origin, then  $D _k $ can be seen as the superposition of a particle of charge +2 ($ D _0 $) and $k $ electrons ($A _0 $). Because of the $\mathbb{Z}  _2  $ identification of eq.~(\ref{z2}), the terms in (\ref{vdk}) corresponding to the $ A _0 $ NUTs come in mirror symmetric pairs, with each pair contributing charge $-1 $. If one pair of NUTs is moved to the origin, we can instead view $D_k $ as the superposition of a proton (AH) and $k-1$ electrons. For both $A _k $ and $D _k $, $Q$  depends only on the asymptotic topology of the manifold, and its value would be the same for a different configuration of NUTs giving the same Chern number. But this is  reasonable: in a system of charged particles, the dominant part of the asymptotic field depends only on the total charge and not on the details of the configuration.

 Let us close this section with a remark on the signature $\tau$. According to the original proposal \cite{Atiyah:2012hw}, $\tau$  gives the baryon number of the manifold. For both $A _k $ and $D _k $, $|\tau| =k$. The fact that for  $k >0 $ the signature of $A _k $ does not agree  with the baryon number of a system of $k + 1 $ electrons suggests to drop the identification of signature with baryon number.

\section{Energy}
We  now discuss how to construct energy functionals. 
While conserved quantum numbers of a particle, like charge, are associated to topological invariants of its modelling manifold, it seems natural to describe dynamical entities, like energy, in terms of geometrical properties of the manifold, like a Riemannian metric or a connection.
A first possibility would be to consider integrals constructed out of the Riemann tensor and its contractions, but the fact that ALF gravitational instantons are Ricci flat and self-dual severely restricts the number of possible constructions, and it does not seem to be possible to get anything but topological quantities via this route. However, ALF gravitational instantons  have non-trivial homology in middle dimension: $H  _2 ( M _k, \mathbb{Z}   ) =\mathbb{Z}  ^k $ where $M _k $ stands for either $A _k $ or $D _k $ \cite{Hausel:2004vv,Etesi:2006bc}. Geometry allows one to select preferred representatives among the 2-cycles generating the homology: those of minimal area. The two energy functionals we propose are built from either the area or the Gaussian curvature of these preferred 2-cycles. We show the construction for  $A _{k-1} $ and $D _k $. We work with $A _{k-1} $ because it has  $k$, rather than $k + 1 $, NUTs and is more easily compared to $D _k $. It is convenient at this point to fix a value for the parameter $\m$. In order for the circles of the asymptotic fibration to have the same length of $2 \pi $ for both $ A _{ k-1 }$ and $D _k $, we choose $\m =1/2 $ for $A _{ k-1 }$ and $\m =1 $ for $D _k $.

Let us start with $A _{ k-1 }$, $k \geq 2 $. There is a basis of $k-1 $ ordered, independent generators of $H _2 (A _{ k-1 }, \mathbb{Z}   )$ that are related to the simple roots of the Lie algebra $A _{ k-1 }$ and intersect according to its Cartan matrix. A very natural way of building representatives of these 2-cycles as  submanifolds of minimal area has been shown in \cite{Sen:1997tw}: consider  the two NUTs $p _i $, $p _{  i + 1 } $ and orient the coordinate axes so that the line between $p _i $ and $p _{i+1} $ lies along the $x$-axis. At each point along this line there is a circle of radius $1/V (x) $ parametrised by $\psi$. Since at the NUTs the radius collapses to zero, the surface defined by the union of all these circles, which we denote by $S _{ i,i + 1 } $, is topologically a 2-sphere. Its (unoriented) area is given by
\begin{equation}
\int _0 ^{ 2 \pi } V  ^{-1} \mathrm{d} \psi  \int _{ x_i } ^{ x_{i + 1 } } V \mathrm{d} x =2 \pi   |x _{i + 1} - x _i  | 
=2 \pi   \norm {p _{i + 1} - p _i  },
\end{equation}
a constant factor times the Euclidean distance between the two NUTs. If instead of a line we  had taken any other curve connecting $p _i $ and $p _{ i + 1 }$, the area would have been the same constant factor times the Euclidean length of the curve, hence the 2-cycle built above is of minimal area. This construction can evidently be done for any pair of distinct NUTs $p_i $, $p _j $, giving the minimal 2-cycle $S _{ i,j } $. While  $\{S _{ 1,2 }\,, \ldots, S _{ k-1,k }\}$ is a basis for $H _2 (A _{ k-1 }, \mathbb{Z}  )$, all pairs of distinct NUTs  play an equal r\^ole in $A _{ k-1 }$ and it would be unnatural to consider only the basis  2-cycles.
The natural choice is to consider instead all the minimal 2-cycles connecting pairs of distinct NUTs. Such a choice is invariant under the Weyl group of the Lie algebra $A _{ k-1 }$, the symmetric group $\mathcal{S} _{ k }$.
 The sets $\{p _i  \} $, $i =1, \ldots, k $, of NUT positions and $\{e _i\}$, $i =1, \ldots ,k $, of canonical basis vectors of $ \mathbb{R}  ^k $ have the same cardinality and so a one-to-one correspondence $p_i \leftrightarrow e _i $ can be established between them. A possible choice for the set of all roots of $A _{ k -1 }$ is $\{\pm (e _i-e _j )$, $i \neq j =1 , \ldots, k\}$. The root $e _i - e _j $ is associated with $S _{ i,j }$ and its negative with the same 2-cycle taken with the opposite orientation, $ S _{ j,i }$. 
Our first proposal for an energy functional is simply to take $\pi$ times the sum of the inverse areas of the minimal 2-cycles $S _{ i,j }$:
\begin{equation}
\label{eak1} 
E ^{ (1) }_{ A _{ k-1 }} = 
\sum _{i< j  =1 }^{k}\frac{1}{\norm {p _i - p _j }}.
\end{equation}

Our second proposal for an energy functional still involves the  2-cycles $S _{ i,j }$. The endpoints of these 2-cycles, the NUTs $p _i $ and $p _j $,  are geometrically preferred points. It is then interesting to calculate the Gaussian curvature $K _{ i,j }$ of $S _{ i,j } $ at the  points $p _i $ and $ p _j $. Without  loss of generality, we can relabel these points so that $i =1 $, $j =2 $ and orient the axes so that $p _1=(x _1, 0, 0) $, $p _2 =(x _2 , 0, 0) $. The metric on $S _{ 1,2 }$ is that of a surface of revolution, $\mathrm{d} s ^2 = V \mathrm{d} x ^2  + V ^{-1}\mathrm{d} \psi ^2 $.
 Its Gaussian curvature $K_{ 1,2 }$ is independent of $\psi$ and is given by, see e.g.~\cite{DoCarmo:1976um},
\begin{equation}
\label{gc} 
K _{ 1,2 }  =- \frac{1}{2} \ \frac{\partial ^2 (V ^{-1} )}{\partial x ^2 }.
\end{equation} 
In calculating (\ref{gc}) at $p _1 $  we take advantage of the fact that many terms are zero. Introduce the shorthand notation $ d _i =\norm {p - p_i } $, $d_0 = \norm{p} $, $d _{ij} =\norm {p _i  - p_j } $ and group the terms in $V ^{-1} $  under a common denominator:
\begin{align}
&V ^{-1} =\frac{f}{f + g }, \qquad  \qquad 
f =\prod   _{ l =1 }^k  d_l, \qquad \qquad  g = \frac{1}{2}  \sum _{l =1 }^k  d_1 \ldots  \widehat{ d _l} \ldots  d _k,
\end{align} 
where terms with a hat are to be omitted. Then 
\begin{equation} 
K_{ 1,2 } =- \frac{1}{2} \frac{1}{(f + g )^3 } \Big( ( f _{ xx }g -f g _{ x x } )(f + g ) -2 (f _x + g _x )(f _x g-f g _x ) \Big).
\end{equation} 
In the limit $p \rightarrow p _1   $ one has 
\begin{equation}
d _l  \rightarrow
\begin{cases} d _{ l 1} &\text{ for $l \neq 1 $}\\
0 &\text{otherwise} \end{cases}, \qquad 
\frac{ \partial d _l}{ \partial x }  \rightarrow
\begin{cases} \frac{ x _1  -x _l }{d _{ l 1}}&\text{ for $l \neq 1 $}\\
$1 $ &\text{otherwise} \end{cases},\qquad 
\frac{\partial ^2  d _l}{\partial x ^2}  \rightarrow
\begin{cases} \frac{y _l ^2  + z _l  ^2}{d _{ l 1} ^3} &\text{ for $l \neq 1 $}\\
0 &\text{otherwise} \end{cases}.
\end{equation} 
Therefore in this limit
\begin{align} 
\nonumber
f &\rightarrow 0, \quad  f _x \rightarrow d_{21} \ldots d_{k1} ,\quad  f _{ x x } \rightarrow \left. 2 (d _{2} \ldots d _{k} )_x  \right |_{ p _1 }, \\
 g &\rightarrow  \frac{1}{2}  d _{21} \ldots d _{k1} , \quad 
 g _x \rightarrow \left.  \frac{1}{2}  (d _{2} \ldots d _{k} )_x \right |_{ p _1 }+  \frac{1}{2} \sum _{ i =2 }^k  d _{21} \ldots \widehat{d _{i1}} \ldots d _{k1} .
\end{align}
  The expression for $K _{ 1,2 }$ simplifies to
\begin{align}
\nonumber
K _{ 1,2 }(p _1 ) &=-\frac{1}{2 g ^2 } \Big( f _{ x x } g  - 2 f _x ( f _x + g _x ) \Big) \\ 
\nonumber&=
- \frac{1}{2} \frac{4}{(d _{21} \ldots d _{k1} )^2 } \Bigg[ 2(d _{21} \ldots d _{k1} )_x \cdot \frac{1}{2} d _{21} \ldots d _{k1} -2 (d _{21} \ldots d _{k1} )^2 
	\\ \nonumber
	&\quad -2 d _{21} \ldots d _{k1} \left( \frac{1}{2} (d _{21} \ldots d _{k1} )_x +\frac{1}{2} \left(  d _{31} \ldots d _{k1} + d _{21} d _{41} \ldots d _{k1} + \ldots \right)  \right) \Bigg] 
	\\ &= 4 \left( 1 + \frac{1}{2 d _{21} } + \frac{1}{2d _{31}  } + \ldots + \frac{1}{2 d _{k1} } \right).
\end{align} 
More generally, for any 2-cycle $S _{ i,j }$,
\begin{equation}
\label{kk} 
K_{ i,j }  (p _i )=4 \left( 1 + \frac{1}{2 d _{1i} } + \frac{1}{2d _{2i}  } + \ldots +\widehat{\frac{1}{2d _{ ii }}} + \ldots +  \frac{1}{2 d _{ki} } \right)
.
\end{equation} 
Note that (\ref{kk})  does not involve the point $p _j $ in any particular way, so that $K _{ i,j }(p_i)=K _{ i,l }(p_i ) $ for any $l \neq i $. Therefore we define $K(p _i ) = K _{ i,l }(p_i ) $  for any 2-cycle $S _{ i,l } $ with $l\neq i $. 
Summing over all the NUTs, dividing by $4$,  and reverting to the original notation  $d_{ij} =\norm{p _i -p _j }$ we get
\begin{equation}
\label{eak2} 
E ^{ (2) }_{ A _{ k-1 }} =
\frac{1}{4}\sum _{ i  =1 }^k K(p _i ) =
k +  \sum _{ i< j =1 }^k \frac{1}{\norm{p _i-p _j }} 
.
\end{equation} 


Expressions (\ref{eak1}) and (\ref{eak2}) are very similar. In both cases the  Coulomb interaction energy of a system of $k$  particles having the same charge is reproduced and the construction, which involves only pairs of electrons, does not run into self-energy problems. The only difference between $E ^{ (1)}_{ A _{ k-1 }}  $ and $E ^{ (2)}_{ A _{ k-1 }} $ is that the latter contains an additive constant equal to the number of electrons.
It seems natural to relate the length of  the asymptotic circles to the classical electron radius $r _e =e ^2 /(m _e c ^2  )$, where $e$ and $m _e  $ are the charge and rest mass of the electron.
Equation (\ref{eak2}) is dimensionless. In dimensional form, with $\eta = r_e/2$, and multiplying by an overall constant factor $m_e c^2$, we get
\begin{equation}
\label{eak2d} 
E ^{ (2) }_{ A _{ k-1 }} =
 k \,m _e c ^2  +  \sum _{ i< j =1 }^k \frac{e ^2 }{\norm{p _i-p _j }},
\end{equation} 
the sum of the rest masses and the interaction energy of $k$ electrons.
  

Consider now the $ D _k $ family, $k \geq 2 $. Since our analysis is based on eq.~(\ref{dkmetric}), which holds only asymptotically, we expect the energy functionals that we derive to be accurate if all the $p _i $ are far from the origin.
 In order to construct $E ^{ (1)}_{ D _k } $, we need to sum the inverse  areas of the 2-cycles corresponding to all the roots of the $ D _k $ Lie algebra. A possible choice of roots is $\{ \pm( e _i \pm e _j  ),\ i<j =1, \ldots , k\} $.\footnote{
The cases $k = 2,3 $ are degenerate. For $k =3 $, the Lie algebra $ D _3 $ is isomorphic to  $A _3 $. The equivalence can be checked by verifying that the Cartan matrix of $D _3 $ associated to the ordered set of simple roots \mbox{$\{ e _2 -e _3,\, e _1 -e _2 ,\, e _2 + e _3 \} $} is equal to the Cartan matrix of $A _3 $. Similarly one can show that  $D _2  $ is isomorphic to $A_1 \times A _1 $.}
As before, we have the correspondence $p _i \leftrightarrow e _i $, hence the 2-cycles  $S _{ \pm i, \pm j } $, connecting the NUTs  $\pm p _i $ to the NUTs  $\pm p _j $, correspond to the roots $\pm e _i \pm e _j $.
 Multiplying by $\pi$ and summing  the inverse areas, we obtain
\begin{equation}
\label{ek1} 
E ^{ (1) }_{ D _k } =\sum _{ i < j =1 } ^k \left(\frac{1}{\norm{p _i -p _j}} +  \frac{1}{ \norm{p_i + p_j }} \right).
\end{equation}
As mentioned in section \ref{charge}, we can relate $D _k $ to either $k-1 $ electrons and a proton or $k$ electrons and a particle of charge +2.
However, eq.~(\ref{ek1}) contains only interaction terms of  electron-electron type, i.e. involving the distances $\norm{p _i \pm p _j }$, $i \neq j $, and no interactions of proton-electron type, which involve the distances $\norm{p _i -0 }$ and should come with the opposite sign.
While it is possible to construct additional 2-cycles corresponding to the proton-electron interactions, it is difficult to get the right numerical factors. Further investigation of the area of these 2-cycles would be worthwhile, but for the moment we opt not to consider $E ^{ (1) }_{ D _k }  $ as a suitable energy functional for $D _k $.

Let us, instead, construct $E ^{ (2) }_{ D _k }$.  As before, the Gaussian curvature of a 2-cycle $S _{ ij }$ can be calculated using eq.~(\ref{gc}). We set $p _{  i + k } =-p _{i }$, $i =1, \ldots , k$, and 
\begin{equation} 
V ^{-1} =\frac{f}{ f + g + h } , \qquad  f=\prod _{ l =0 } ^{ 2k } d_l , \qquad g =-\frac{f}{d_0 }  , \qquad  h =\frac{1}{4} \sum _{ i =1 }^{2k} d _0 \,d _1 \ldots \widehat{d _i }\ldots d _{ 2k } .
\end{equation}
Proceeding similarly as we did for $A _{ k-1 }$, we find that the Gaussian curvature of the
 2-cycle $S _{ i,j } $, $i \neq j$, $i \neq j +k$, $j \neq i + k $, in the limit $p\rightarrow p _i $ is
\begin{align}
\label{kdk} 
K _{ i,j } (p _i ) &=
4 \left( 1- \frac{2}{\norm{p _i}} + \frac{1}{2}  \sum _{l =1 ,\,l\neq i}^k \left( \frac{1}{\norm{p _l -p _i }} +
 \frac{1}{\norm{p _l +p _i }} \right) \right).
\end{align}
In order to get (\ref{kdk}) we have used $d _{ ij }=\norm{p _i - p _j}$ if either $j, i<k$ or $j, i>k$, $d _{ ij }=\norm{p _i + p _j}$ otherwise, and $d _{ 0j }=\norm{p _j}$.
Eq.~(\ref{kdk}) does  not involve $p _j $ in any particular way, so we can again define $K(p _i )= K _{ i,j }(p _i ) $ for any 2-cycle $S _{ i,j }$ with $i \neq j $, $ i\neq j +k $, $j \neq i +k  $. Note that $K(p _i ) =K(p _{i+k} ) $.
Since $p _i $ is identified with $p _{ i + k } $, we sum $K (p _i ) /4$ over $p _i $ for $i =1,\, \ldots,\, k $ only, obtaining
\begin{equation}
\label{ek3} 
\begin{split} 
E^{ (2) }_{ D _k } &=
\frac{1}{4} \sum _{ i =1 } ^{ k }K( p _i ) =
 k- \sum _{ i =1 } ^{ k } \frac{2}{\norm{p _i} } + \sum _{ i< j =1} ^{ k } \left( \frac{1}{\norm{p _i- p _j }}+\frac{1}{\norm{p _i + p _j} } \right) .
\end{split} 
\end{equation} 
If $\norm{p _i -p _j }\ll\norm{p _i }$, $\norm{p _j } $,  the term $1/\norm{p _i + p _j } $ is negligible, and (\ref{ek3}) reduces to an additive constant, equal to the number of electrons, plus the Coulomb interaction energy of $k$ electrons and a particle of charge $+ 2 $.
 If all the $p _i $ are far from the origin, corrections to $E ^{ (2) }_{ D _k } $ due to the different behaviour of the exact metric near the origin are small  and we expect them to be related to the rest mass of the positively charged particle. 
In dimensional form, with $\eta = r_e$, and multiplying by an overall constant factor $m_e c^2$,
\begin{equation}
E^{ (2) }_{ D _k } =
 k\,m _e c ^2 - \sum _{ i =1 } ^{ k } \frac{2 e ^2 }{\norm{p _i} } + \sum _{ i< j =1} ^{ k } \left( \frac{e ^2 }{\norm{p _i- p _j }}+\frac{e ^2 }{\norm{p _i + p _j} } \right) .
\end{equation}

Let us now see what happens to  $E ^{ (2) }_{ D _k }$ in the limit $p _k \rightarrow 0 $. We cannot take this limit directly in (\ref{ek3}), but   we can calculate the  Gaussian curvature of the 2-cycles $S _{ \pm i,\pm j } $ using the metric (\ref{dkmetric}) with $V$ as in (\ref{vdkbis}). Nothing particular happens to the 2-cycles $\{S _{ \pm i, \pm j },i,j =1, \ldots , k-1, i\neq j   \}$. The 2-cycles 
$\{ S _{ \pm i, \pm k } , i =1, \ldots, k-1 \}$ now connect the points $\{\pm p_ i\} $ to the origin.  Their geometry near the origin is not described accurately by the metric (\ref{dkmetric}), but  the description is accurate near  $p _i $. 
The Gaussian  curvature at the point $p _i \neq 0$ of any 2-cycle $S _{i, j } $, $i\neq j $, $i \neq j +k $, $j \neq i + k $, calculated using  eq.~(\ref{gc}), with $V$ as in (\ref{vdkbis}), is
\begin{equation}
K ( p _i ) =4 \left( 1- \frac{1}{\norm{p _i }}+ \frac{1}{2} \sum _{ l =1,\, l\neq i }^{ k-1 } \left( \frac{1}{\norm{p _l - p _i }} +  \frac{1}{\norm{p _l +  p _i }} \right) \right).
\end{equation} 
Therefore, neglecting the contribution of the origin, we have
\begin{equation}
\label{ek4}
\begin{split} 
\left. E   ^{ (2) }_{ D _k } \right |_{ p _k =0 }&=
\frac{1}{4}\sum _{ i =1 }^{ k-1 } K  ( p _i ) =
k-1- \sum _{ i =1 }^ {k-1 }\frac{1}{\norm{p _i }}+  \sum _{ i<j =1}^{ k-1 } \left( \frac{1}{\norm{p _i - p _j }} +  \frac{1}{\norm{p _i +  p _j }} \right) .
\end{split} 
\end{equation} 
In dimensional form,
\begin{equation}
\label{ek4d}
\left. E   ^{ (2) }_{ D _k } \right |_{ p _k =0 }=
 (k-1) m _e c ^2 - \sum _{ i =1 }^ {k-1 }\frac{e ^2 }{\norm{p _i }}+  \sum _{ i<j =1}^{ k-1 } \left( \frac{e ^2 }{\norm{p _i - p _j }} +  \frac{e ^2 }{\norm{p _i +  p _j }} \right) .
\end{equation} 
We expect the contribution of the origin to modify the rest mass term; the interaction terms in (\ref{ek4d}) can be recognised as the Coulomb interaction energy of a proton and $k-1 $ electrons.

\section{Conclusions}
In \cite{Atiyah:2012hw},  self-dual 4-dimensional manifolds were used in an attempt to provide a geometrical description of matter. Their various topologies are potentially related to the particle quantum numbers, like baryon and lepton number, and electric charge. Their geometry is potentially related to particle energies and interactions. 
Many issues were left open, for example dynamics was not considered and the number of concrete examples proposed was relatively small.

In this paper we have started filling some of the gaps. The class of concrete examples has been extended by  two infinite families, the $A _k $ and $D _k $ ALF gravitational instantons. The former has been proposed as a model for $k + 1$ electrons, while the latter for a positively charged particle and some electrons. The positively charged particle can be a proton, in which case there are $k-1$ electrons, or a particle of charge $+ 2 $, in which case there are $k$ electrons. 
As a first step towards dynamics we have constructed energy functionals which reproduce the appropriate Coulomb interaction energy for the particle systems considered.
Our formulae are precise for the $A _k $ manifolds, whose metric is known exactly, but only approximate for the $D _k $ manifolds, whose metric is only accurately known away from the origin.
%

Our constructions  have hinted at the important r\^{o}le of 2-dimensional substructures in this geometrical approach. Finally, this work supports the identification of charge with the first Chern number of the asymptotic fibration but suggests a rethinking of the identification of baryon number with signature.

\acknowledgments
G.F.~acknowledges funding from a Marie Curie Actions ESR grant and thanks Sergey Cherkis and Bernd Schroers for useful and interesting discussions.
\bibliographystyle{jhep}
\bibliography{Geom_models}
\end{document}